\documentclass[a4paper]{jpconf}
\usepackage{graphicx}
\usepackage{bm}
\usepackage{amsmath}
\usepackage{citesort}
\begin{document}
\title{Effect of Fermi surface evolution on superconducting gap in superconducting topological insulator}

\author{Tatsuki Hashimoto, Keiji Yada, Ai Yamakage, Masatoshi Sato and Yukio Tanaka}
\address{Depertment of Applide Physics, Nagoya University, Nagoya 464-8603, Japan}

\ead{hashimoto@rover.nuap.nagoya-u.ac.jp}

\begin{abstract}
We study bulk electronic states of superconducting topological insulator, which  is the promising candidate for topological superconductor. 
Recent experiments suggest that the three-dimensional Fermi surface evolves into two-dimensional one. 
We show that the superconducting energy gap structure on the Fermi surface systematically changes with this evolution. 
It is clarified that the bulk electronic properties such as spin-lattice relaxation rate and specific heat depend on the shape of the Fermi surface and the type of the energy gap function. 
These results serve as a guide to determine the pairing symmetry of Cu$_x$Bi$_2$Se$_3$.
\end{abstract}

\section{Introduction}
 Recently, topological insulator(TI) and topological superconductor(TSC) 
become star materials drawing intense research interest because of their novel physical properties and potential applications in electronic devices\cite{Hasan,XLQi,Tanaka1,Ando}. 
The remarkable feature of TI is the presence of 
surface Dirac cone protected by the topological invariant $Z_{2}$ 
defined in bulk time-reversal invariant insulating state\cite{XLQi,Ando}. 
Thanks to the existence of the surface Dirac cone, 
we can expect new exotic transport phenomena and future application to spintronics devices in TI. 
The superconducting analogue of topological insulator is dubbed as TSC, which also has gappless surface state called 
surface Andreev bound state(SABS) protected by the 
bulk topological invariant \cite{Schnyder,Sato,Sato10,Tanaka1}. 
One of the important feature of TSC is that 
the SABS can be Majorana fermion\cite{Wilczek}. 
There are several proposals that 
Majorana fermion is applicable to fault tolerant quantum computation\cite{Nayak}.

One of the typical example of the three-dimensional TIs is Bi$_2$Se$_3$\cite{Xia,HZhang}, which is widely studied since it has a simple Dirac-cone at $\Gamma$ point and large bulk band gap comparable to the room temperature\cite{Ando}. 
Recent intense researches of Bi$_2$Se$_3$ reveal that Cu doped Bi$_2$Se$_3$ shows superconductivity with transition temperature $T_c=3.8$K\cite{Hor}. Soon after the observation of superconductivity, angle resolved photoemission spectroscopy measurement in the normal state has clarifid that the topological surface state remains even after the Cu doping \cite{Wray1}. In this sence, Cu$_x$Bi$_2$Se$_3$ can be called superconducting topological insulator(STI) and considered as a best platform for studying the relation between TI and superconductivity. Moreover, owing to its peculiar band structure from the strong spin-orbit interaction, we can naturally imagine that TSC can be realized.

Tunneling spectroscopy measurement is one of the most promising tool to detect the SABS, which can be a clue to reveal the pairing symmetry. 
Actually, the SABSs in the cuprates and Sr$_2$RuO$_4$ manifest as a zero bias conductance peak (ZBCP)
 \cite{Tanaka2,Kashiwaya1,Yamashiro1,Yamashiro2,Kashiwaya2,Mao}, 
and the presence of ZBCP becomes an evidence for unconventional pairing like 
$d$-wave or $p$-wave.  
There are some experimental results that indicate the realization of topological superconductivity in Cu$_x$Bi$_2$Se$_3$. 
ZBCP has been reported in the point contact spectroscopy by Sasaki $et$ $al$\cite{Sasaki}. 
By the help of theoretical calculations for possible pair potentials 
\cite{Sasaki,Yamakage1,Takami,HaoLee,Hsieh}, it has been revealed that the 
resulting ZBCP originates from SABS. 
After the first observation of ZBCP, 
other groups confirmed the existence of the 
ZBCP by point contact spectroscopy \cite{Kirzhner,Chen}. 
In addition to the above surface sensitive measurements, 
bulk physical quantities, $e.g.$, specific heat\cite{Kriener1}, superfluid density\cite{Kriener2}, and upper critical field\cite{Bay} have been measured. In Ref.\cite{Kriener2} anomalous behavior of the superfluid density has been reported, which indicates possible realization of unconventional superconductivity in  Cu$_x$Bi$_2$Se$_3$. Moreover, upper critical field data by Bay $et$ $al$. have suggested spin-triplet pairing \cite{Bay}.

There are some conflicting experimental results that indicate 
the non-topological superconductivity without SABS in Cu$_x$Bi$_2$Se$_3$. 
In the scanning tunneling spectroscopy, the ZBCP has not been observed, but $U$-shaped fully-gapped spectrum has appeared\cite{Levy,Peng}. 
Nevertheless, more recently, Mizushima $et$ $al.$ theoretically has shown that simple $U$-shaped spectra can not be realized by non-topological pairing 
due to the inevitable mixing of pair potentials with different symmetry 
on the surface \cite{Mizushima}. 
Since the pairing symmetry of this material has not been clarified yet, it is a hot topic to reveal the unconventional superconductivity\cite{FuBerg,Hashimoto,Yip,Nagai,Nagai1,Nagai2,Zocher,MichaeliFu}.

Although it has been considered that the Fermi surface of Cu$_x$Bi$_2$Se$_3$ is three-demensional, most recently, Lahoud $et$ $al.$ has revealed that the Fermi surface can be changed into quasi-two-dimensional one depending on Cu concentration\cite{Lahoud}.
Thus, it is really interesting and timely to study how the energy gap on the Fermi surface and resulting various physical quantities change with the evolution of the Fermi surface. To clarify these properties are really useful to identify the pairing symmetry and establish the topological superconductivity in Cu$_x$Bi$_2$Se$_3$. 
It is noted that the $U$-shaped spectra can be interpreted by the odd-parity pairing if we introduce the evolution of the Fermi surface from spheroidal to cyrindirical shape \cite{Mizushima}. 

In this paper, we show the change of the superconducting energy gap structure on the Fermi surface with the evolution of the shape of the Fermi surface. 
We calculate the density of state (DOS), 
temperature dependence of the longitudinal spin-lattice relaxation rate in NMR 
and the specific heat for STI assuming possible pair potentials. 

This paper is organized as follows. In Sec.\ref{sec_model} we present the model Hamiltonian for STI. We also show the analytical formulae of the superconducting gap functions.
In Sec.\ref{sec_DOS}, we show the changes of the energy gap on the Fermi surface and the DOS with evolution of the Fermi surfece. Next, we calculate the temperature dependence of the spin-lattice relaxation rate in Sec.\ref{sec_NMR} and the specific heat in Sec.\ref{sec_spe}. Finally, we summarize this paper in Sec.\ref{sec_summary}.
\section{Model}\label{sec_model}
\par
To describe electronic states of STI,
we start with the Bogoliubov-de Gennes Hamiltonian proposed in Ref.\cite{FuBerg}, 
\begin{eqnarray}
H({\bm k})=H_0({\bm k})\tau_z+\hat{\Delta}_i\tau_x,
\label{hamiltonian}
\end{eqnarray}
where $\tau$ is the Pauli matrix in the Nambu space and $i(=1,2,3,4)$ denotes the type of pair potential.
$H_0({\bm k})$ is the normal part of the Hamiltonian given by
\begin{eqnarray}
H_0({\bm k})&=&c({\bm k})+m({\bm k})\sigma_x+v_z\sin k_z\sigma_y +v(\sin k_xs_y-\sin k_ys_x)\sigma_z,
\label{h00} \\ 
m({\bm k})&=&m_0+2m_1(1-\cos k_z)+2m_2(2-\cos k_x-\cos k_y),
\label{m} \\
c({\bm k})&=&-\mu+2c_1(1-\cos k_z)+2c_2(2-\cos k_x-\cos k_y).
\label{c}
\end{eqnarray}
$s_i$ and $\sigma_i$ $(i=x,y,z)$ are the Pauli matrices in the spin and orbital space, respectively\cite{Liu}.
In this model, we consider the cubic lattice structure while the crystal structure of Cu$_x$Bi$_2$Se$_3$ is rhonbohedral one\cite{MaoYamakage}.
This simplification does not affect the low energy electronic structures.
The basis of the orbitals in Eq.(\ref{h00}) consists of two effective $p_z$ orbitals of Se and Bi on the upper and lower sides in a quintuple layer.
Hereafter, we call this basis the ``orbital basis".
On the other hand, we refer to the labels after the diagonalization of $H_0(\bm k)$ as the ``band basis",
\begin{align}
U^\dagger({\bm k}) H_0({\bm k})U({\bm k})=c({\bm k})\tilde{\sigma}_0s_0+\eta({\bm k})\tilde{\sigma}_zs_0,\label{bandbasis}
\end{align}
where $\eta({\bm k})=\sqrt{m^2({\bm k})+v^2(\sin^2k_xa+\sin^2k_ya) + v_z^2\sin^2k_zc}$ and $\tilde{\sigma}_z$ is the Pauli matrix which denotes the band index, i.e., $\tilde{\sigma}_z=1$ for the conduction band and $\tilde{\sigma}_z=-1$ for the valence band.

Before we move to the description of the types of the pair potential,
we discuss the shape of the Fermi surface. 
If we adopt the parameters of the normal-state Hamiltonian as reported in Ref.\cite{Sasaki}, 
it gives a three-dimensional Fermi surface around $\Gamma$-point.
These parameters well describe the electronic structure of Bi$_2$Se$_3$.
However, in Cu$_x$Bi$_2$Se$_3$, since Cu is intercalated between the quintuple layers, 
the values of the hopping terms along the $z$-direction ($c_1$, $m_1$, $v_z$) and the chemical potential $\mu$ can be modulated due to the change of the $c$-axis length and the electron doping, respectively.
In this paper, we use the parameter reported in Ref.\cite{Sasaki} for $\mu=0.40$. For $\mu=0.65$, we use the half value of $c_1$, $m_1$ and $v_z$. For other $\mu$, we interpolate the value of $c_1$, $m_1$ and $v_z$ linearly.
For larger $\mu$ and/or smaller values of hopping terms along the $z$-axis ($c_1$, $m_1$, $v_z$),
the shape of the Fermi surface becomes cylindrical.

Next, we explain the possible types of the pair potential for this two orbital model\cite{FuBerg}.
Since Cu$_x$Bi$_2$Se$_3$ is not a strongly correlated system,
we assume that the Cooper pairs within a unit cell are dominant and therefore
the pair potential does not depend on ${\bm k}$.
In this case, Fermi statistics allows six types of pair potential: $\Delta \sigma_0 s_0$, $\Delta \sigma_x s_0$, $\Delta \sigma_y s_z$, $\Delta \sigma_z s_0$, $\Delta \sigma_y s_x$ and $\Delta \sigma_y s_y$.
They are classified into four types of irreducible representation of the $D_{3d}$ point group,
$A_{1g}$ ($\Delta \sigma_0 s_0$ and $\Delta \sigma_x s_0$), $A_{1u}$ ($\Delta \sigma_y s_z$), $A_{2u}$ ($\Delta \sigma_z s_0$), and $E_{u}$ ($\Delta \sigma_y s_x$, $\Delta \sigma_y s_y$).
In this paper, we consider the cases of $\hat{\Delta}_{1}=\Delta \sigma_0 s_0$, $\hat{\Delta}_{2}=\Delta \sigma_y s_z$, $\hat{\Delta}_{3}=\Delta \sigma_z s_0$ and $\hat{\Delta}_{4}=\Delta \sigma_y s_x$.
We do not take into account of $\Delta \sigma_x s_0$ since spin-orbit interaction in Eq.(\ref{h00}) does not favor $\Delta \sigma_x s_0$ pairing.
Note that the result for $\Delta\sigma_ys_y$ is obtained by that of $\Delta_4$ by four-fold rotation around the z-axis.
In the above pairings, $\hat{\Delta}_{1}$ is an even-parity BCS pairing and $\hat{\Delta}_{2}$, $\hat{\Delta}_{3}$ and $\hat{\Delta}_{4}$ are odd-parity pairings.

Now we consider the pair potential in the band basis as introduced in Eq. (\ref{bandbasis})
by the unitary transformation,
\begin{align}
\Delta\sigma_i s_j
\rightarrow
U^\dagger({\bm k})\Delta\sigma_is_jU({\bm k})
=
\Delta\tilde{\sigma}_i s_j.
\end{align}
Here, we consider the electron doped TI as realized in Cu$_x$Bi$_2$Se$_3$, where the Fermi surface consists of only the conduction band.
Since of band gap is much larger than the superconducting gap\cite{Wray1},
we can ignore the inter-band pairing and the intra-band pairing within the valence band.
Then, we obtain the superconducting gap in the tight-binding model from the $(1,1)$ component of the Puli matrix $\tilde{\sigma}_i$ in the band basis. 
\begin{align}
|\Delta_1({\bm k})|
 &=\Delta\label{gap1}\\
|\Delta_2({\bm k})|
 &=\Delta\sqrt{\frac{m({\bm k})^2}{m({\bm k})^2+v_z^2 \sin^2k_zc}\frac{v^2(\sin^2k_xa+\sin^2k_ya)}{\eta({\bm k})}+\frac{v_z^2 \sin^2k_zc}{m^2({\bm k})+v_z^2 \sin^2k_zc}}\label{gap2}\\
|\Delta_3({\bm k})|
 &=\Delta\sqrt{\frac{v^2(\sin^2k_xa+\sin^2k_ya)}{\eta^2({\bm k})}}\label{gap3}\\
|\Delta_4({\bm k})|
 &=\Delta\sqrt{\frac{v^2\sin^2k_xa+v_z^2\sin^2k_zc}{\eta^2({\bm k})}}\label{gap4},
\end{align}
Detailed derivations of the pair potentials in the band basis have been reported in Ref.\cite{Hashimoto,Yip}.
\section{Energy gap structure with the evolution of Fermi surface}\label{sec_DOS}
The energy gap structures for three-dimensional Fermi surface have been clarified in the previous studies. 
In this case, $\Delta_1$ and $\Delta_2$ show fully-gapped structures while $\Delta_3$ and $\Delta_4$ have point nodes along the $k_z$- and $k_y$-axis, respectively\cite{FuBerg,HaoLee,Sasaki}.
In the following, we show the change of the energy gap structures on the Fermi surface by Cu intercalation.
We also show the DOS which is useful to understand the calculated results of the spin-lattice relaxation rate and the specific heat
shown in the later sections.

 In the calculation of the DOS, we adopt the quasiclassical approximation $\Delta\ll v_Fk_F$ where $v_F$ is the group velocity along the normal vector to the Fermi surface.
Then, the DOS in the superconducting state $N_s(E)$ normalized by its value in the normal state $N_n(E=0)$ is given by
\begin{align}
\frac{N_s(E)}{N_n(0)}=&\int_{\rm{FS}}{\rm{Re}} {\left[\frac{|E|}{v_F\sqrt{E^2 - \Delta_i^2({\bm k})}}\right] dS}/\int_{\rm{FS}}\frac{1}{v_F}dS\nonumber\\
\simeq&\int_{\rm{FS}}{\rm{Re}} {\left[\frac{|E|}{\sqrt{E^2 - \Delta_i^2({\bm k})}}\right] dS},\label{DOS}
\end{align}
where $\int_{\rm{FS}}dS$ denotes the integral over the Fermi surfece.
In the second line in Eq.(\ref{DOS}),
we assume that $v_F$ is constant on the Fermi surface for simplicity.
Since the ${\bm k}$-dependence of the energy gap and the DOS are trivial for $\Delta_1$, we show the DOS for $\Delta_2$, $\Delta_3$ and $\Delta_4$.
\subsection{Energy gap structure and the DOS for $\Delta_2$}
\begin{figure}[tbp]
\begin{center}
\includegraphics[width=14cm]{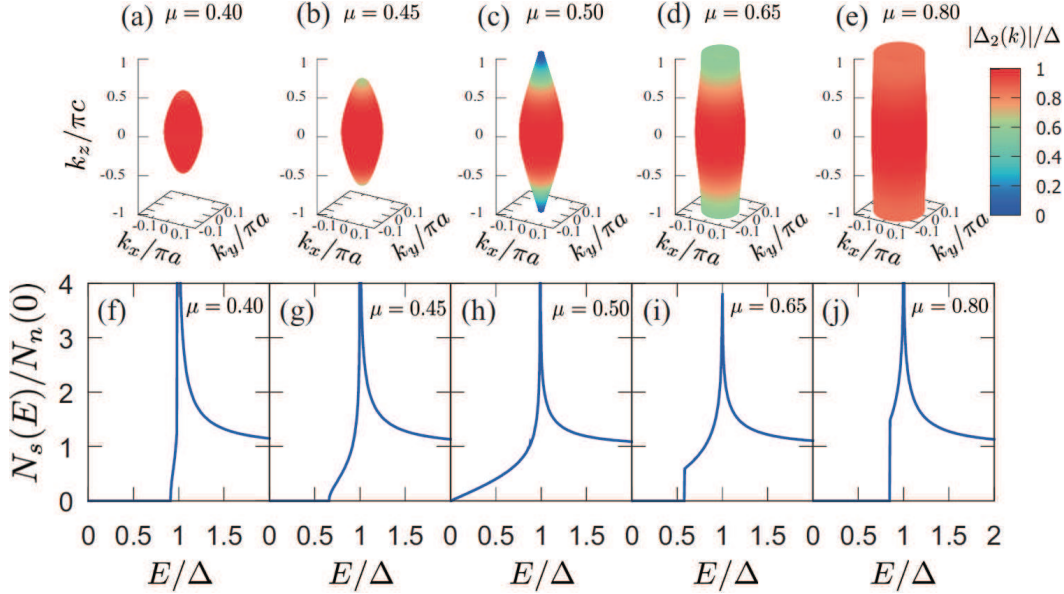}
\caption{(a)-(e)Evolution of the Fermi surface and superconducting gap, (f)-(i)the density of state, as a function of the chemical potential $\mu$ in the case of $\Delta_2$. The color of the Fermi surface indicates the magnitude of the superconducting gap at the Fermi momenta.}\label{DOS_d2}
\end{center}
\end{figure}
In Fig.\ref{DOS_d3}, we show the evolution of the Fermi surface and the superconducting gap (a)-(e) and the DOS (f)-(j) in the case of $\Delta_2$.
The color on the Fermi surface indicates the magnitude of the energy gap at the Fermi momenta. 
In the case of $\mu=0.40$ and $0.45$, the Fermi surface is three-dimensional(Fig.\ref{DOS_d2}(a) and (b)).
With increasing $\mu$, the volume of the Fermi surface becomes larger and the top and the bottom of the Fermi surface reach the boundary of the Brillouin zone at $\mu=0.50$ as seen from Fig.\ref{DOS_d2}(c).
With increasing $\mu$, the Fermi surface becomes cylindrical and the two-dimensionality of the Fermi surface is much stronger for larger $\mu$(Fig.\ref{DOS_d2}(d) and (e)).
As seen from Fig.\ref{DOS_d2}, the energy gap other than the case of $\mu=0.50$ have fully-gapped structures.
On the other hand, it is closed at $(k_x,k_y,k_z)=(0,0,\pm\pi/c)$ in the case of $\mu=0.50$.
It is noted that the topological phase changes between the spherical and the cylindrical Fermi surfaces.
In the fully-gapped odd-parity superconductors with time-reversal symmetry,
the number of the time-reversal-invariant momenta enclosed by the Fermi surface is odd (even), the three-dimensional winding number is odd (even)\cite{Sato}. In the present cases, the number of the time-reversal-invariant momenta enclosed by the Fermi surface is one for $\mu<0.50$ and two for $\mu>0.50$,
and therefore, the topological nature of the STI for $\Delta_2$ changes\cite{Lahoud}.
Since the topological phase does not change without an energy gap closing,
nodal point appears between two different topological phases at $\mu=0.50$.
Owing to this gap closing,
the energy gap near $(k_x,k_y,k_z)=(0,0,\pm\pi/c)$ is suppressed.
Thus, $\Delta_2({\bm k})$ is highly anisotropic for $\mu=0.45$ and $0.65$ where the ratio of the maximum to the minimum of the energy gap is almost 2.
On the other hand, anisotropy is small for $\mu=0.40$ and $0.80$.
Then, the DOS for $\mu=0.40$ and $0.80$ has a large peak at $E/\Delta=1$ and rapidly decreases to zero for $E/\Delta<1$.
This is similar to that for the isotropic-gap superconductor like $\Delta_1$.
\subsection{Energy gap structure and the DOS for $\Delta_3$}
\begin{figure}[tbp]
\begin{center}
\includegraphics[width=14cm]{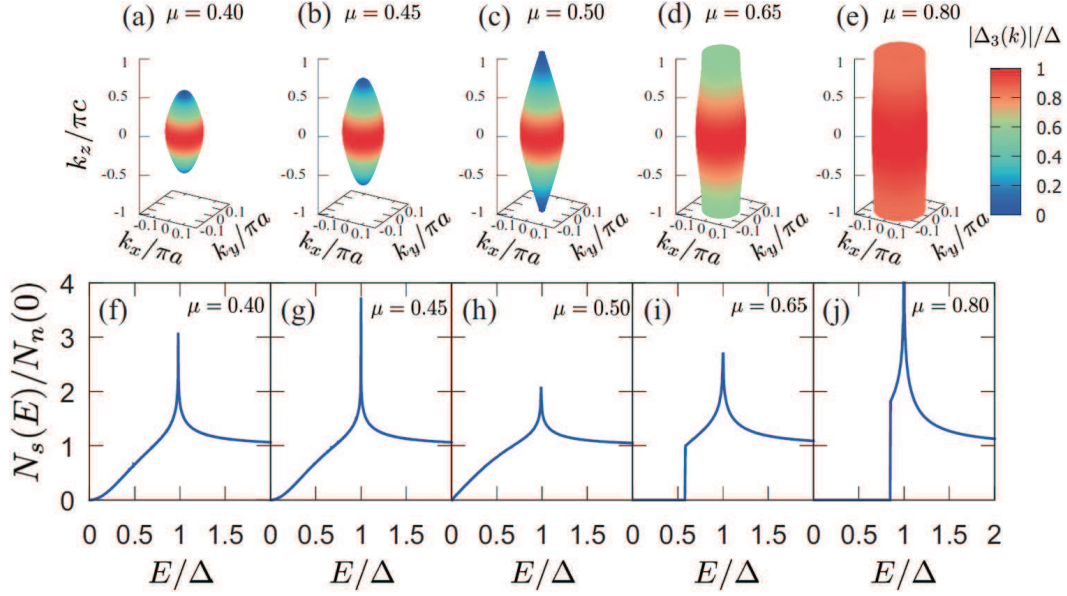}
\caption{(a)-(e)Evolution of the Fermi surface and superconducting gap, (f)-(i)the density of state, as a function of the chemical potential $\mu$ in the case of $\Delta_3$. The color of the Fermi surface indicates the magnitude of the superconducting gap at the Fermi momenta.}\label{DOS_d3}
\end{center}
\end{figure}
In Fig. \ref{DOS_d3}, we show the evolution of the Fermi surface and the superconducting gap(a)-(e) and the DOS(f)-(j) in the case of $\Delta_3$.
As seen from Eq. (\ref{gap3}), $\Delta_3({\bm k})=0$ along the $k_z$-axis ($k_x=k_y=0$).
Thus, in the case of the three-dimensional Fermi surface ($\mu<0.50$), the point nodes appear at the north and south poles on the Fermi surface as seen from Figs. \ref{DOS_d3}(a) and (b).
Due to these point nodes, the DOS near $E=0$ is proportional to $E^2$ as shown in Figs. \ref{DOS_d3}(f) and (g).
Since the Fermi surface is vertically long, the curvature of the Fermi surface has a maximum at nodal points.
This curvature diverges at $\mu=0.50$ where the Fermi surface reaches the zone boundary $k_z=\pi$.
Then, the coefficient of $E^2$ in the DOS diverges and $E$-linear behavior appears as shown in Fig. \ref{DOS_d3}(h).
In contrast, in the case of two-dimensional Fermi surface $\mu>0.50$, the DOS shows fully-gapped structures since the nodal points disappear.
With increasing $\mu$, the anisotropy of the energy gap becomes small since the two-dimensionality of the Fermi surface increases.
\subsection{Energy gap structure and the DOS for $\Delta_4$}
\begin{figure}[tbp]
\begin{center}
\includegraphics[width=14cm]{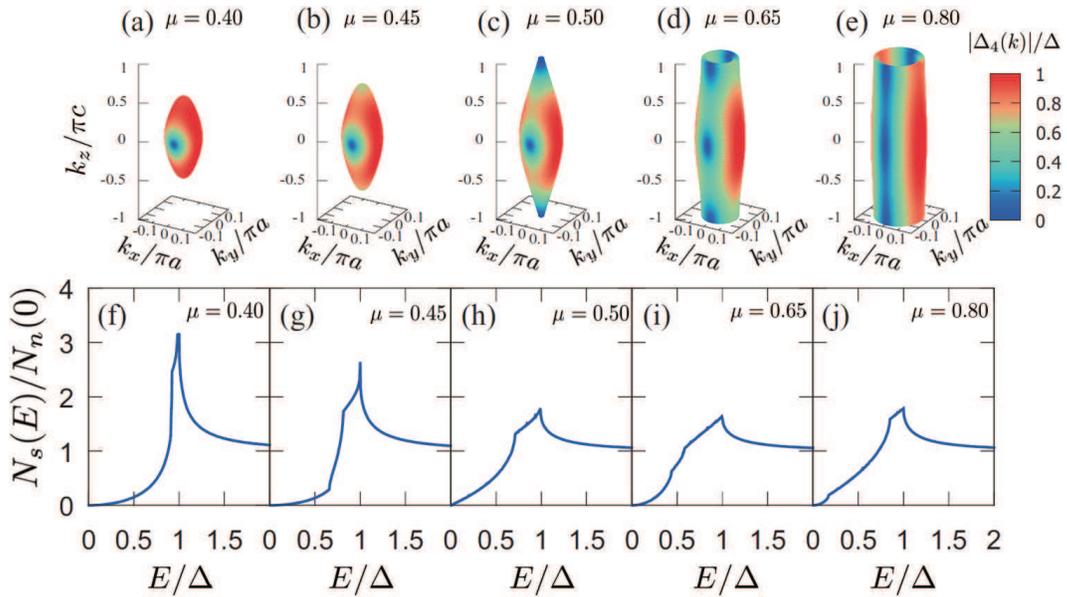}
\caption{(a)-(e)Evolution of the Fermi surface and superconducting gap, (f)-(i)the density of state, as a function of the chemical potential $\mu$ in the case of $\Delta_4$. The color of the Fermi surface indicates the magnitude of the superconducting gap at the Fermi momenta.}\label{DOS_d4}
\end{center}
\end{figure}
In Fig.\ref{DOS_d4}, we show the evolution of the Fermi surface and the superconducting gap(a)-(e) and the DOS(f)-(j) in the case of $\Delta_4$.
In the case of the three-dimensional Fermi surface,
point nodes exist along the $k_y$-axis ($k_x=k_z=0$).
Thus, the DOS near $E=0$ is proportional to $E^2$ as shown in Figs. \ref{DOS_d3}(f) and (g).
As compared to the case of $\Delta_3$, the coefficient of $E^2$ in the DOS is smaller since the curvature of the Fermi surface at the point nodes is smaller than that for $\Delta_3$.
When the Fermi surface becomes two-dimensional by increasing $\mu$,
additional point nodes appear at $(k_x, k_z)=(0, \pi)$ as seen from Figs.\ref{DOS_d4}(d) and (e).
Since the number of the point nodes becomes twice,
the coefficient of $E^2$ in the DOS becomes larger than that for the three-dimensional Fermi surface $\mu<0.50$.
In addition, due to the in-plane anisotropy of the superconducting gap in Eq. (\ref{gap4}),
the magnitude of the energy gap for $k_x=0$ is much smaller than that for $k_y=0$.
It has been clarified that line nodes which connect above point nodes appears in the $k_yk_z$-plane in two-dimensional limit ($v_z=0$).
It is noted that $E$-linear behavior appears at $\mu=0.50$ by the reason similar to the case of $\Delta_3$.
\section{Spin-lattice relaxation rate}\label{sec_NMR}
The temperature dependence of the spin-lattice relaxation rate in superconducting states has a peak just below $T_c$,
which is called Hebel-Slichter peak\cite{HebelSlichter}.
The Hebel-Slichter peak originates from the peak of DOS near $E=\Delta$
and, in general, it can be suppressed by the anisotropy of the superconducting gap, quasi-particle damping and inhomogeneities.
The temperature dependence of the spin-lattice relaxation rate near $T=0$ also has characteristic behavior depending on the gap structure\cite{SigristUeda}.
In the case of fully gapped superconductors, 
the spin-lattice relaxation rate shows an exponential behavior at low temperature.
On the other hand, that for the superconductors with point nodes or line nodes shows a power-law behavior reflecting the DOS near $E=0$.
In this section, we calculate the temperature dependence of the spin-lattice relaxation rate for the STI with the possible pair potentials in each case of Fermi surface. 
 Furthermore, we reveal the effect of quasi-particle damping.
We compare the height of the Hebel-Slichter peak for the STI with that for the isotropic-gap superconductor by changing the magnitude of the quasi-particle damping.

The normalized spin-lattice relaxation rate $T_1^{-1}$ is given by 
\begin{align}
\frac{T_{1N}}{T_1}=\frac{2}{N_n(0)^2}\int^{\infty}_{0}dE N_s^2(E)
\left(
1+\frac{<\Delta_{i\uparrow\downarrow}>}{E}
\frac{<\Delta^*_{i\uparrow\downarrow}>}{E}
\right)
\left(
-\frac{\partial f(E)}{\partial E}
\right),\label{t1}
\end{align}
where $N_s(E)$ is the DOS in the superconducting state given by Eq. (\ref{DOS}),  $f(E)$ is the Fermi distribution function and $<\Delta_{i\uparrow\downarrow}>$ $(i=1,2,3,4)$ denotes the average of the pair potential on the Fermi surface\cite{SigristUeda}.
This average part equal to zero for $\Delta_2$, $\Delta_3$ and $\Delta_4$ because these pair potentials are odd parity. 
For the temperature dependence of pair potential, we consider a phenomenological form obtained from the BCS theory\cite{BM},
\begin{eqnarray}
\Delta(T)=\alpha k_BT_c\tanh\left(1.74\sqrt{\frac{T_c}{T}-1}\right),\label{tdel}
\end{eqnarray}
where $\alpha$ is a coupling constant.
In the conventional BCS superconductor, $\alpha$ is known to be $1.76$. 
However, $\alpha$ offten deviates from the BCS value\cite{Padamsee}.
The estimated $\alpha$ for Cu$_x$Bi$_2$Se$_3$ is $2.3$ from the measurements of upper critical magnetic field in Ref.\cite{Kriener1}.
Thus, we use $\alpha = 2.3$ in this paper.
To consider the quasi-particle damping effect,
we add the imaginary part $E$ by replacing $E$ to $E+i\delta$ in Eq.(\ref{DOS}),
where $\delta$ denotes the magnitude of the quasi-particle damping.
In this paper we set the normalized magnitude of the quasi-particle damping as $\delta/\Delta=10^{-3}$, $10^{-2}$ and $10^{-1}$.

\begin{figure}[tbp]
\begin{center}
\includegraphics[width=4cm]{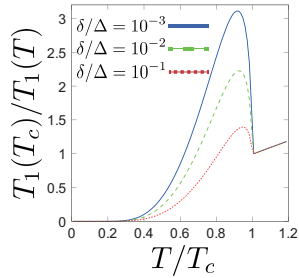}
\caption{Temperature dependence of the spin-lattice relaxation rate $T_1^{-1}$ in the case of isotropic gap pairing $\Delta_1$. The normalized magnitude of the quasi-particle damping is set to $\delta/\Delta=10^{-3}$(blue solid line), $10^{-3}$(green dashed line) and $10^{-1}$(red dotted line)}\label{T1_d1}
\end{center}
\end{figure}

In Fig.\ref{T1_d1}, we show the temperature dependence of the spin-lattice relaxation rate for the isotoropic-gap superconductor $\Delta_1$ with $\delta/\Delta=10^{-3}$, $10^{-2}$ and $10^{-1}$. 
We can clearly see a prominent Hebel-Slichter peak.
The height of the Hebel-Slichter peak decreases with increasing $\delta/\Delta$.
However, this peak still remains in the case with large quasi-particle damping $\delta/\Delta=10^{-1}$.
Since the energy gap structure for $\Delta_1$ is unchanged by the concentration of Cu, 
the same results are obtained for both two-dimensional Fermi surface and three-dimensional one.

\begin{figure}[htbp]
\begin{center}
\includegraphics[width=15cm]{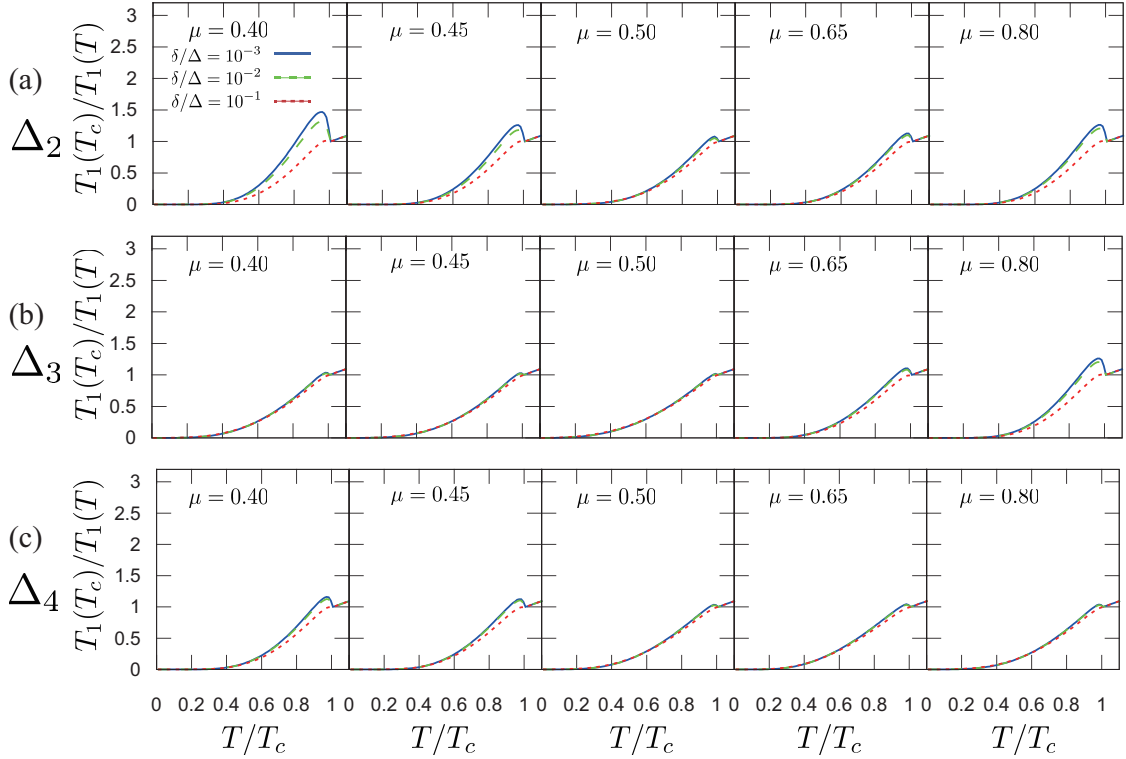}
\caption{Temperature dependence of $T_1(T_c)/T_1(T)$ for (a)$\Delta_2$, (b)$\Delta_3$ and (c)$\Delta_4$ with $\mu=0.40$, $0.45$, $0.50$, $0.65$ and $0.80$. The normalized magnitude of the quasi-particle damping is set to $\delta/\Delta=10^{-3}$(blue solid line), $10^{-3}$(green dashed line) and $10^{-1}$(red dotted line)}\label{T1_fig}
\end{center}
\end{figure}

Next, we show the temperature dependence of the spin-lattice relaxation rate $T_1(T_c)/T_1(T)$ for $\Delta_2$, $\Delta_3$ and $\Delta_4$ in Figs. \ref{T1_fig}(a), (b) and (c), respectively.
In the case of $\Delta_2$, the DOS has a Hebel-Slichter peak near for $\delta/\Delta=10^{-3}$.
The height of this peak becomes smallest at $\mu=0.50$ where the energy gap is the most anisotropic due to the topological phase transition.
In addition, the height of this peak is smaller than that for $\Delta_1$.
Thus, we can not clearly see the Hebel-Slichter peak for $\delta/\Delta=10^{-1}$.

In the case of $\Delta_3$, the Hebel-Slichter peak for three-dimensional Fermi surface ($\mu<0.50$)
is quite suppressed 
even in the case of small quasi-particle damping $\delta/\Delta=10^{-3}$ as shown in Fig.\ref{T1_fig}(b).
This is because the DOS at $E/\Delta=1$ is smaller than that of the other pairings
due to the formation of point nodes.
By contrast, with increasing $\mu$, the energy spectra become fully-gapped.
As a result, the Hebel Slichter peak appears for two-dimensional Fermi surface ($\mu<0.50$) with small quasi-particle damping ($\delta/\Delta=10^{-3}$ and $10^{-2}$).
Due to the disappearance of point nodes in two-dimensional Fermi surface,
the power-law behavior of $T_1(T_c)/T_1(T)$ at low temperature for three-dimensional Fermi surface
changes to exponential one.

In the case of $\Delta_4$, the height of the Hebel-Slichter peak decreases with the evolution of the Fermi surface.
This is consistent with the suppression of the peak in the DOS at $E/\Delta=1$.
In other words, the suppression of the Hebel-Slichter peak is owing to the emergence of the additional point nodes at zone boundary.
Note that the Hebel-Slichter peak does not appear for large magnitude of the quasi-particle damping in both cases with two-dimensional and three-dimensional Fermi surface.

Here, we summarize the results of $T_1$.
In the weak quasi-particle damping cases ($\delta/\Delta=10^{-3}$ and $10^{-2}$),
the temperature dependence of $T_1(T_c)/T_1(T)$ is different for each type of pairing.
The Hebel-Slichter peak appears for (i) $\Delta_1$, (ii) $\Delta_2$, (iii) $\Delta_3$ with two-dimensional Fermi surface or (iv) $\Delta_4$ with three-dimensional Fermi surface. 
The height of the Hebel-Slichter peak for $\Delta_1$ is much larger than that for other pairings. 
Thus, with the strong quasi-particle damping ($\delta/\Delta=10^{-1}$),
the Hebel-Slichter peaks for all odd-parity pairings ($\Delta_2$, $\Delta_3$ and $\Delta_4$) disappear
while that for even-parity pairings ($\Delta_1$) remains.
In usual cases, since the magnitude of pair potential is much smaller than the chemical potential,
the quasi-particle damping effect is strong.
Therefore, it is concluded that the absence of the Hebel-Slichter peak has a possibility supporting the odd-parity superconductivity.

\section{Specific heat}\label{sec_spe}
\begin{figure}[tbp]
\begin{center}
\includegraphics[width=15cm]{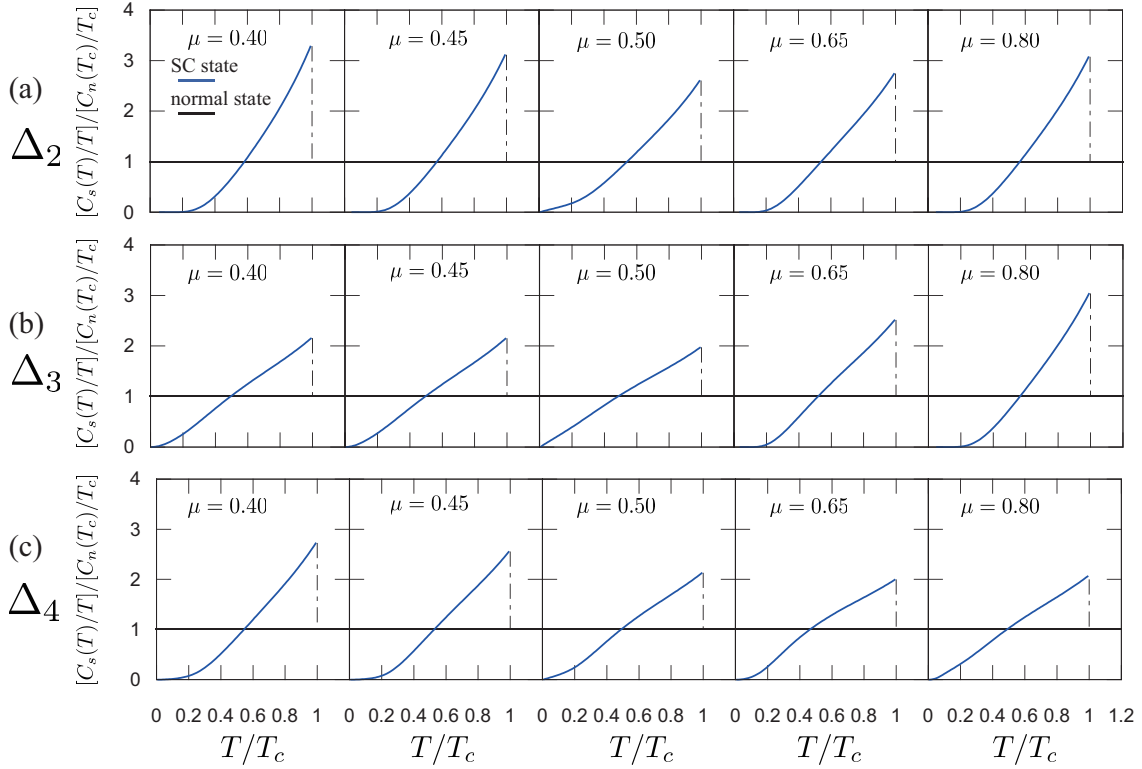}
\caption{Temperature dependence of $C_{s}(T)/T$ for (a)$\Delta_2$, (b)$\Delta_3$ and (c)$\Delta_4$ with $\mu=0.40$, $0.45$, $0.50$, $0.65$ and $0.80$. The blue(black) line shows the specific heat in the superconducting(normal) state.}\label{spe_fig}
\end{center}
\end{figure}
In this section, we calculate the temperature dependence of the specific heat of the STI for the possible pair potentials.
The specific heat is given by
\begin{eqnarray}
C_{s}=\frac{2\beta}{N}\sum_{{\bm k}}\left(E_{i}^2({{\bm k}})+\beta E_{i}({{\bm k}})\frac{\partial \Delta}{\partial \beta}\frac{\partial E_{i}({{\bm k}})}{\partial \Delta}\right)\left(-\frac{\partial f(E_i({\bm k}))}{\partial E_i({\bm k})}\right),\label{spe_equ}
\end{eqnarray}
where $N$ is the number of unit cells, $\beta=1/k_BT$ and $E_i({\bm k})$ $(i=2,3,4)$ is the energy dispersion for each possible pair potential.
The detailed forms of $E_i({\bm k})$ is given in Ref.\cite{Hashimoto}.
For the temperature dependence of the pair potential,
we use Eq. (\ref{tdel}) with $\alpha=2.3$ which is the same as that used in the calculation of spin-lattice relaxation rate.
In general, the temperature dependence of $C_{s}(T)/T$ for a full gap superconductor has an exponential behavior at low temperature.
In the case of superconductors with point nodes and line nodes, $C_{s}(T)/T$ is proportional to $T^2$ and $T$ near $T=0$, respectively.
On the other hand, 
the magnitude of a jump of $C_{s}(T)/T$ at $T=T_c$ increases with the coupling constant $\alpha$.
In addition, since the following entropy valance relation must be satisfied,
\begin{align}
\int_0^{T_c}dT\frac{C_s(T)-C_n(T)}{T}=0,
\end{align}
the specific heat at low temperatures and that at high temperatures more or less influence each other.

In Fig.\ref{spe_fig}, we show the specific heat for the STI as a function of the temperature. The first row shows the temperature dependence of $C_{s}(T)/T$ for $\Delta_2$ with $\mu=0.40$, $0.45$, $0.50$, $0.65$ and $0.80$. The second and third row shows that for $\Delta_3$ and $ \Delta_4$ respectively.
In the case of $\Delta_2$, with the evolution of the Fermi surface, the energy gap closes at $\mu=0.50$. Near the transition point, the magnitude of the specific heat jump is the smallest among the calculated results for $\Delta_2$ since the anisotropy of the energy gap becomes the largest. 
The magnitude of the specific heat jump at $T=T_c$ becomes larger as $\mu$ is apart from $\mu=0.50$.
At low temperature, $C_{s}(T)/T$ has an exponential behavior except for $\mu=0.50$ as shown Fig. \ref{spe_fig}(a). For $\mu=0.50$, it is proportional to $T$, which is consistent with $E$-linear behavior in the DOS.

As we can see in Fig. \ref{spe_fig}(b), the temperature dependence of $C_{s}(T)/T$ for $\Delta_3$ changes with the evolution of the Fermi surface.
This is because the energy gap structure changes from point nodal one to fully-gapped one with the increase of $\mu$. 
In the same reason, $C_{s}(T)/T$ near $T=0$ shows a $T^2$ behavior and an exponential behavior for three-dimensional and two-dimensional Fermi surfaces, respectively.
The magnitude of the specific heat normalized by its value in the normal state just below $T_c$ is almost two for three-dimensional Fermi surface.
For $\mu>0.50$, this value increase with $\mu$ due to the suppression of anisotropy of the energy gap.

According to Fig.\ref{spe_fig}(c), the temperature dependence of $C_{s}(T)/T$ for $\Delta_4$ also changes with the evolution of the Fermi surface. 
In the case of $\Delta_4$, since the energy spectra have point nodes both for two-dimensional and three-dimensional Fermi surface, $C_{s}(T)/T$ is proportional to $T^2$.
However, the magnitude of the specific heat jump with three-dimensional Fermi surface is larger than that for two-dimensional one.
This is because the number of point nodes increases in two-dimensional Fermi surface.

Here, we compare the calculated results with experimental ones measured by Kriener $et$ $al$\cite{Kriener1}. 
In their measurements, the ratio of the specific heat in the superconducting state to that in the normal state is about three. 
It is consistent with our results for the case of (i)$\Delta_2$, (ii)$\Delta_3$ with two-dimensional Fermi surface or (iii) $\Delta_4$ with three-dimensional Fermi surface. 

Before we move to the summary of this section, we refer to the recently discovered STI, Cu$_x$(PbSe)$_5$(Bi$_2$Se$_3$)$_6$ (CPSBS)\cite{CPSBS}. In CPSBS, the experimental results of the specific heat show a nodal behavior\cite{CPSBS}. 
They are consistent with our calculated results for the case of (i)$\Delta_3$ with three-dimensional Fermi surface or (ii)$\Delta_4$ with two-dimensional Fermi surface. In addition, there is a sufficient possibility that two-dimensional Fermi surface is realized in CPSBS since it is suggested that the parent material (PbSe)$_5$(Bi$_2$Se$_3$)$_6$ has two-dimensional Fermi surface\cite{PSBS,CPSBS}. From this point of view, the realization of $\Delta_4$ is most likely. However,  since the electronic states of this material are not sufficiently clear at present, further experimental and theoretical studies are required to reveal the pairing symmetry.

In conclusion, 
we have shown that the wide variation of the temperature dependence of the specific heat appears depending on the shape of the Fermi surface. 
Especially, the magnitude of the specific heat jump changes systematically with the evolution of the Fermi surface. 
In the case of $\Delta_2$, the magnitude of the specific heat jump once decreases, up to $\mu=0.50$, and increases with increasing $\mu$. On the other hand, the magnitude of the specific heat jump for $\Delta_3$($\Delta_4$) is enhanced(reduced) almost monotonically with increasing $\mu$. 
Namely, the systematical changes of the specific heat jump are different in each pair potential. 
For this reason, it is possible to identify the pairing symmetry from the measurements of specific heat by changing the amount of Cu intercalation. 
\section{Summary}\label{sec_summary}
In this paper, we have studied 
bulk electronic states of superconducting topological insulator, 
Cu$_{x}$Bi$_{2}$Se$_{3}$.  
We have focused on the evolution of the 
Fermi surface changing from three-dimensional to two-dimensional shape and its infuence on bulk physical quantities. 
It has been shown that the superconducting energy gap structure on the Fermi surface systematically changes with this evolution. 
We have clarified that the spin-lattice relaxation rate and the specific heat depend on the shape of the Fermi surface and the type of the energy gap function. 
These obtained results are useful to identify the pairing symmetry of Cu$_x$Bi$_2$Se$_3$. 

In this paper, we have studied four types of pair potential generated by local attractive interaction. As a future problem, it is interesting to extend our work for anisotropic pairing realized by correlation effect\cite{HaoWang,ChenWan}.
\section{Acknowledgements}\label{sec_Ack}
We are grateful to S. Onari, S. Sasaki, M. Kriener, K. Segawa and Y. Ando for valuable discussions. 
This work was supported by the "Topological Quantum Phenomena" (No. 22103002) Grant-in Aid for Scientific Research on Innovative Areas from the Ministry of Education, Culture, Sports, Science and Technology (MEXT) of Japan, Grant-in-aid for JSPS Fellows (No. 26010542) (T.H.) and  Grant-in-Aid for Scientific Research B (No. 25287085) (M.S).
\section*{References}

\bibliography{iopart-num}

\end{document}